\documentclass[intlimits,twoside,a4paper]{article}

\usepackage[cp1251]{inputenc}

\usepackage[eqsecnum]{cmpj3}

\issue{2021}{24}{4}{43501}
\doinumber{10.5488/CMP.24.43501}
\title[Time series analysis of friction force]%
{Time series analysis of friction force at self-affine mode of ice surface softening%
}
\author[A. V. Khomenko, D. T. Logvinenko]{A. V. Khomenko\orcid{0000-0001-8755-9592}\refaddr{label1,label2}\thanks{Corresponding author: \email{o.khomenko@mss.sumdu.edu.ua}.},
        D. T. Logvinenko\orcid{0000-0003-1719-9223}\refaddr{label1}}
\addresses{
\addr{label1} Sumy State University, 2 Rimsky-Korsakov St., 40007 Sumy, Ukraine
\addr{label2} Peter Gr\"unberg Institut-1, Forschungszentrum-J\"ulich, D-52425 J\"ulich, Germany
}


\date{Received April 23, 2021, in final form July 7, 2021}
\begin{document}

\maketitle

\begin{abstract}
The self-affine mode of ice softening during friction is investigated within the rheological model for viscoelastic medium approximation. The different modes of ice rubbing, determined by formation of surface liquid-like layer, are studied. The analysis of time series of friction force is carried out, namely Fourier analysis, construction of autocorrelation and difference autocorrelation functions. The spectral power law is detected for modes of crystalline ice as well as of a mixture of stable ice and metastable softening. The self-similarity and aperiodic character of corresponding time series of friction force are  proved.
%
%
\printkeywords
%
\end{abstract}

\section{Introduction}\label{sec:level1}

Ice friction is an important fundamental and applied problem, involving processes such as creep, fracture and melting \cite{review,Heor_RSA_18,Buzz_JAM_18,Pers_JCP_18,TL_Pers_2016,Pers_15}. The solution of this problem is directly related to the definition of models to describe these processes, the description of the modes which arise while ice sliding, and their properties. In particular, the motion on ice may lead to the melting or softening of the ice surface by friction~\cite{TL_Pers_2016,Pers_15}. It was assumed that the surface layer of the ice that softened during friction is characterized by a high density of defects. Such property resembles the behavior of ultrathin films of a lubricant confined between solid surfaces \cite{Khomenko2010_jfrw,Khomenko2007165,Khom_TEPL2012}. Being two-dimensional systems, they differ in properties from ordinary bulk liquids and may behave abnormally. Accordingly, a nonlinear model of a viscoelastic medium with thermal conductivity has been developed to describe the ice surface premelting within the framework of the theory of phase transitions \cite{ice}. The system of kinetic equations was suggested representing the mutually coordinated behavior of shear strain $\varepsilon$ and stress $\sigma$, and temperature $T$ in the surface ice layer during friction.  Experimental data and calculations \cite{Farad_Disc_2012,phil_mag_A_2000,swed_09,JGRB:JGRB17369,Sukhorukov20131,Klapproth2016169,jgs_2005} have shown that intermittent (stick-slip) friction can be a stochastic regime in which the static and kinetic friction forces change randomly in time. Therefore, the previously obtained transition model was supplemented by taking into account Gaussian fluctuations that play the role of external forces and lead to stick-slip friction \cite{Khomenko2017_TL}. In the basic equations, the intensities of shear strain $\varepsilon$ and stress $\sigma$, and temperature $T$ noises $I_\varepsilon$, $I_\sigma$, and $I_T$ were introduced.

Moreover, in the study \cite{Persson2014} it was shown that the intermittent regime is also self-affine for most natural and practically important surfaces. A new stochastic model has helped to find out the reasons for the appearance of this mode and its features \cite{Khomenko2018_TL}. We used the measure units
\begin{eqnarray}
&& \varepsilon_{s}= \sigma_{s}/G_\varepsilon,
\quad \sigma_{s}=\left(c_p\eta_\varepsilon T_{c}/\tau _{T}\right)^{1/2},
\quad T_{c}, \nonumber \\ && \left( \varepsilon_{s} / \tau_\varepsilon \right)^2, \quad\quad\quad\quad \sigma_{s}^{2}, \quad\quad\quad\quad\quad\quad T_c^2,
\label{1aa} \end{eqnarray}
for magnitudes $\varepsilon $, $\sigma$, $T$, $I_\varepsilon$, $I_\sigma$, and $I_T$, respectively, where $G_{\varepsilon} \equiv \eta_{\varepsilon}/\tau_{\varepsilon}\equiv G(\omega)|_{\omega\to 0}$ is the relaxed value of the ice shear modulus ($\omega$ is the circular frequency of a periodic external influence), $\eta_\varepsilon$ is the effective value of shear viscosity $\eta$, $\tau_{\varepsilon}$ is the relaxation time of ice strain, $c_p$ is the heat capacity, $T_{c}$ is the characteristic temperature, $\tau_{T}\equiv l^2 c_p/\kappa$ is the time of heat conductivity, $l$ is the distance into which heat penetrates the ice, $\kappa$ is the thermal conductivity. The nonlinear relaxation term of strain and feedbacks were considered in the governing equations with the fractional exponent $0<a<1$ \cite{Khomenko2018_TL}:
\begin{eqnarray}
&&\tau_{\varepsilon }\dot{\varepsilon}=-\varepsilon^a + \sigma + \sqrt{I_\varepsilon}\xi_1(t), \label{eq2} \\
&&\tau_{\sigma}\dot{\sigma}=-\sigma +g(T-1)\varepsilon^a +\sqrt{I_\sigma}\xi_2(t), \label{eq3} \\
&&\tau _{T}\dot{T}=(\tau _{T}Q-T) - \sigma \varepsilon^a +\sqrt{I_T}\xi_3(t). \label{eq4}
\end{eqnarray}
Here, the stress relaxation time $\tau_{\sigma}$, the constant $g={G_0}/{G_\varepsilon}<1$ ($G_{0} \equiv G(T=2T_{c})$ is a typical value of modulus) and the thermostat temperature (the background sliding block temperature) $T_e = \tau_{T}Q$ ($Q$~is a heat flow from the sliding block to the surface film) are presented. The function $\xi_i(t)$ describes $\delta$-correlated Gaussian source (white noise) \cite{gardiner2009} with moments:
\begin{equation}
\langle\xi_i(t)\rangle = 0,\quad \langle\xi_i(t)\xi_j(t')\rangle = 2\delta_{ij}\delta(t-t'),
\label{xi_corr}
\end{equation}
where $\langle \ldots \rangle$ denotes a statistical ensemble average value, $\delta(x)$ is the Dirac delta function and $\delta_{ij}$ stands for the Kronecker delta symbol. The relaxation pattern of a viscoelastic ice during friction is construed by the Kelvin-Voigt equation~(\ref{eq2}) ($a=1$), which is commonly recognized in the rheology of ice friction~\cite{Khomenko2010_jfrw,Khomenko2007165,book_Kozin}. The Landau-Khalatnikov-type relaxation equation for stress~(\ref{eq3}) \cite{Khomenko2010_jfrw,Khomenko2007165,kin} takes into account the dependence of the shear modulus on the dimensionless temperature $G(T) = G_{0}(T-1)$. According to our approach  \cite{ice}, equation (\ref{eq3}) is obtained from the theory of phase transitions rather than from rheology. On the other hand, replacing $\varepsilon / \tau_{\sigma}$ by $\partial \varepsilon / \partial t$ in equation~(\ref{eq3}) at $G(T) = {\rm const}$ and $a=1$ we reduce it to the Maxwell-type equation for a viscoelastic medium that is widely approved for  description of ice dynamics \cite{book_Kozin,Persson_book}. Equation~(\ref{eq4}) is the heat conductivity expression specifying the thermal transmission from thermostat to the ice surface, the dissipative heating of the stress-induced viscous flow and a heat source due to dilatation and deformation energy \cite{ice}. The classification of different regions of the phase diagram is based on the following property. When shear strain $\varepsilon=0$, the ice is not premelted. The case $\varepsilon\neq 0$ meets its softening. The $\varepsilon\neq 0$ appears at temperature $T_e$ above critical $T_{c0}=1+g^{-1}$, but at small $T_e\leqslant T_{c0}$, the ice solidifies \cite{ice,Khomenko2017_TL,Khomenko2018_TL}.

It is believed that this paper will provide an opportunity to go forward in establishing the empirical laws of ice rubbing. We are aiming at investigating the time series of friction force to find their frequency and correlational characteristics.

\section{Langevin and Fokker-Planck equations}\label{sec:level2}

From the system (\ref{eq2})--(\ref{eq4}), the Langevin equation is acquired within the framework of the adiabatic approximation $\tau_{\sigma}, \tau_{T} \ll \tau _{\varepsilon}$ in \cite{Khomenko2018_TL}
\begin{equation} \tau_{\varepsilon}\dot\varepsilon = f_a(\varepsilon)+\sqrt{I_a(\varepsilon)}~\xi(t),  \label{VI.5_17a}
\end{equation}
where the deterministic force and the effective noise intensity read
\begin{equation}
f_a (\varepsilon) = - \varepsilon^a + g\varepsilon^a \left(T_e -1 \right) d_a(\varepsilon),~  d_a(\varepsilon){\equiv} (1+g\varepsilon^{2a})^{-1}, \label{X19}
\end{equation}
\begin{equation}
I_a(\varepsilon)\equiv I_\varepsilon + \left[I_\sigma + I_T(g\varepsilon^a)^2\right] d_a^2(\varepsilon).
\label{X3}
\end{equation}

Since Langevin equation is a stochastic differential equation, let us write the corresponding Fokker-Planck equation in the Stratonovich form \cite{Khomenko2018_TL}:
\begin{eqnarray} \label{X5}
\frac{\partial P_a(\varepsilon,t)}{\partial t} =
-\frac{\partial}{\partial\varepsilon}\left[f_a(\varepsilon) P_a(\varepsilon,t)\right] + \frac{\partial}{\partial\varepsilon}\left[
\sqrt{I_a(\varepsilon)}\frac{\partial}{\partial\varepsilon}\sqrt{I_a(\varepsilon)} P_a(\varepsilon,t)\right].
\end{eqnarray}
The stationary ($\partial P_a(\varepsilon,t) / \partial t = 0$) probability density of equation~(\ref{VI.5_17a}) solutions is derived from equation~(\ref{X5}) \cite{Landau}:
\begin{equation}
P_a(\varepsilon)=Z^{-1}\exp\lbrace-U_a(\varepsilon)\rbrace,
\label{a}
\end{equation}
where $Z$ is a normalization constant and the effective potential reads
\begin{equation}
U_a(\varepsilon)=\frac{1}{2}\ln I_a(\varepsilon)-\int\limits^\varepsilon_0{f_a(\varepsilon')\over I_a(\varepsilon')} \rd\varepsilon'. 
\label{VI.5_17}
\end{equation}

The subsequent analysis consists in a numerical solution of the Langevin stochastic differential equation~(\ref{VI.5_17a}) by the Euler method \cite{Khomenko2018_TL,Khomenko2010_jfrw,Khomenko_UJP2009,Khomenko_TechPhys2007}. With this aim, the Stratonovich's stochastic differential equation~(\ref{VI.5_17a}) is transformed into the Ito's one:
\begin{equation}
\rd\varepsilon = \left[f_a(\varepsilon){+} \sqrt{I_a(\varepsilon)}\frac{\partial}{\partial\varepsilon}\sqrt{I_a(\varepsilon)}\right]\rd t {+} \sqrt{I_a(\varepsilon)}\rd W(t), \label{ito} \end{equation}
where $\rd W(t)$ is the Wiener process \cite{gardiner2009}. The determination of the discrete analog of stochastic force differential $\rd W(t) \equiv \sqrt{\triangle t}W_i$, equations (\ref{VI.5_17a}) and (\ref{X3}) lead to the iteration procedure:
\begin{eqnarray}
\varepsilon_{i+1} {=} \varepsilon_i {+} \left(f_a(\varepsilon_i){+}\frac{ga\varepsilon_i^{2a-1}[g I_T(1{-}g\varepsilon_i^{2a}){-}2I_\sigma]}
{(1{+}g\varepsilon_i^{2a})^3}\right)\triangle t + \sqrt{I_a(\varepsilon_i)\triangle t}W_i.
\label{iter}
\end{eqnarray}
Solution of (\ref{iter}) is carried out on the time range $t\in[0,T_m]$ with increment $\triangle t = T_m/N$ ($N$ is the number of iterations). The stochastic force $W_i$ with properties \cite{gardiner2009}
\begin{equation}
\langle W_i \rangle=0, \quad \langle W_i W_{i'} \rangle = 0, \quad \langle W^2_i \rangle \to 2
\label{Sila_diskret}
\end{equation}
can be written using the Box-Muller model \cite{press2007numerical}:
\begin{equation}
W_i = \sqrt{\kappa^2}\sqrt{-2\ln r_1}\cos(2\piup r_2), \quad r_n \in (0,1].
\end{equation}
Here, $\kappa^2=2$ is the variance, $W_i$ is the numbers with properties (\ref{Sila_diskret}) and $r_1$, $r_2$ are pseudo-random numbers characterized by homogenous and periodical distribution.

\section{Analysis of friction force time series}\label{sec:level3}

\begin{figure}[htbp]
\centerline{
\includegraphics[width=0.45\textwidth]{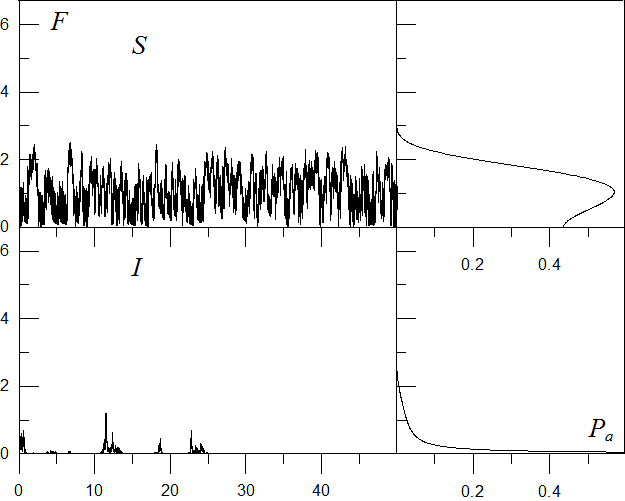} \includegraphics[width=0.45\textwidth]{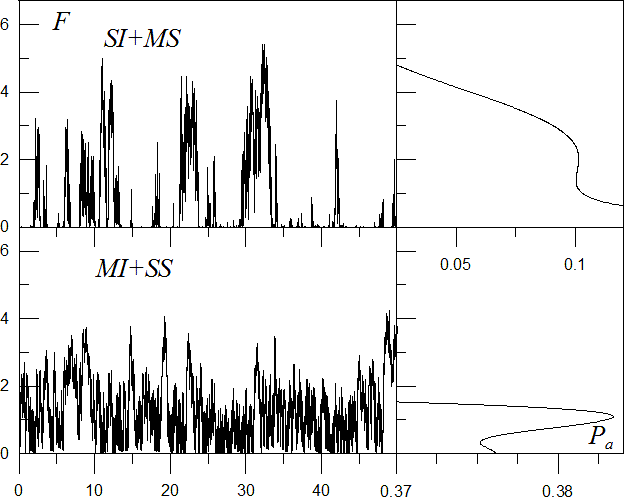}
}
\caption{Friction force time series $F(t)$ (in arbitrary units) and its distribution $P_a$ obtained by solution of (\ref{iter}) and (\ref{a}) ($g = 0.8, T_e = 1.2, a = 0.9, N = 10^6, T_m=50, dt= 5 \cdot 10^{-5}$). The panels meet the parameters:
$I_\sigma = 4.25, I_T = 2$ [softened ice $(S)$];
$I_\sigma = 0, I_T = 5$ [crystalline ice $(I)$];
$I_\sigma = 0, I_T = 55$ [stable ice and metastable softening $(SI+MS)$];
$I_\sigma = 5, I_T = 12$ [metastable ice and stable softening $(MI+SS)$].}
\label{fig1}
\end{figure}

In figure~\ref{fig1}, the time dependencies of frictional force are depicted, which are built on the basis of approximation $F(t) = A G_\varepsilon |\varepsilon(t)|$ in accordance with~\cite{Khomenko2018_TL,Khomenko2017_TL}. Therein, we have shown that $F(t)$ corresponds to the experiments described in  \cite{Farad_Disc_2012,phil_mag_A_2000,swed_09,JGRB:JGRB17369,Sukhorukov20131,Klapproth2016169,jgs_2005} for the following quantities:
1) the strain relaxation time $\tau_\varepsilon \approx 0.1 - 5$~s;
2) the contact area $A\approx 10^{-6} - 10^{-1}$~m$^2$ that is typical of friction of a polymer (poly(methyl methacrylate) PMMA), rubber and steel on ice \cite{Farad_Disc_2012,Klapproth2016169,jgs_2005} as well as for polycrystalline freshwater and saline ice over itself \cite{phil_mag_A_2000,swed_09,JGRB:JGRB17369,Sukhorukov20131};
3) the shear modulus of ice $G_\varepsilon\approx 0.1 - 10$~MPa, i.e., yield stress \cite{Farad_Disc_2012}.
Probability density (\ref{a}) shown in right-hand parts of figure~\ref{fig1} specifies the friction modes represented by left-hand sides of this figure. Panel, where only one maximum of (\ref{a}) exists at most probable $F_0\ne 0$ is related to the softened ice friction ($S$). It is assumed that the total friction force $F$ involves the static and kinetic components. Ice friction region ($\it I$) of the phase diagram is characterized by a single maximum of probability density at $F_0=0$. Two-phase region $\it SI+MS$ is defined by coexistence of distribution maxima $P_a(\varepsilon)$ at zero and nonzero $F_0$ corresponding to the stable ice and its metastable premelting. Probability density $P_a(\varepsilon)$ in the area $\it MI+SS$ has two maxima at $F_0 = 0$ and $F_0\ne 0$ and construes a stick-slip mode when transitions between metastable ice and stable softened ice are realized. It has been revealed that the crystalline ice and $SI+MS$ friction may proceed in the self-affine regime \cite{Khomenko2018_TL}. The existence of heterogeneous quasi-liquid layers containing two phases was confirmed experimentally~\cite{Sazaki24012012}.

\begin{figure}[htbp]
\centerline{
\includegraphics[width=0.5\textwidth]{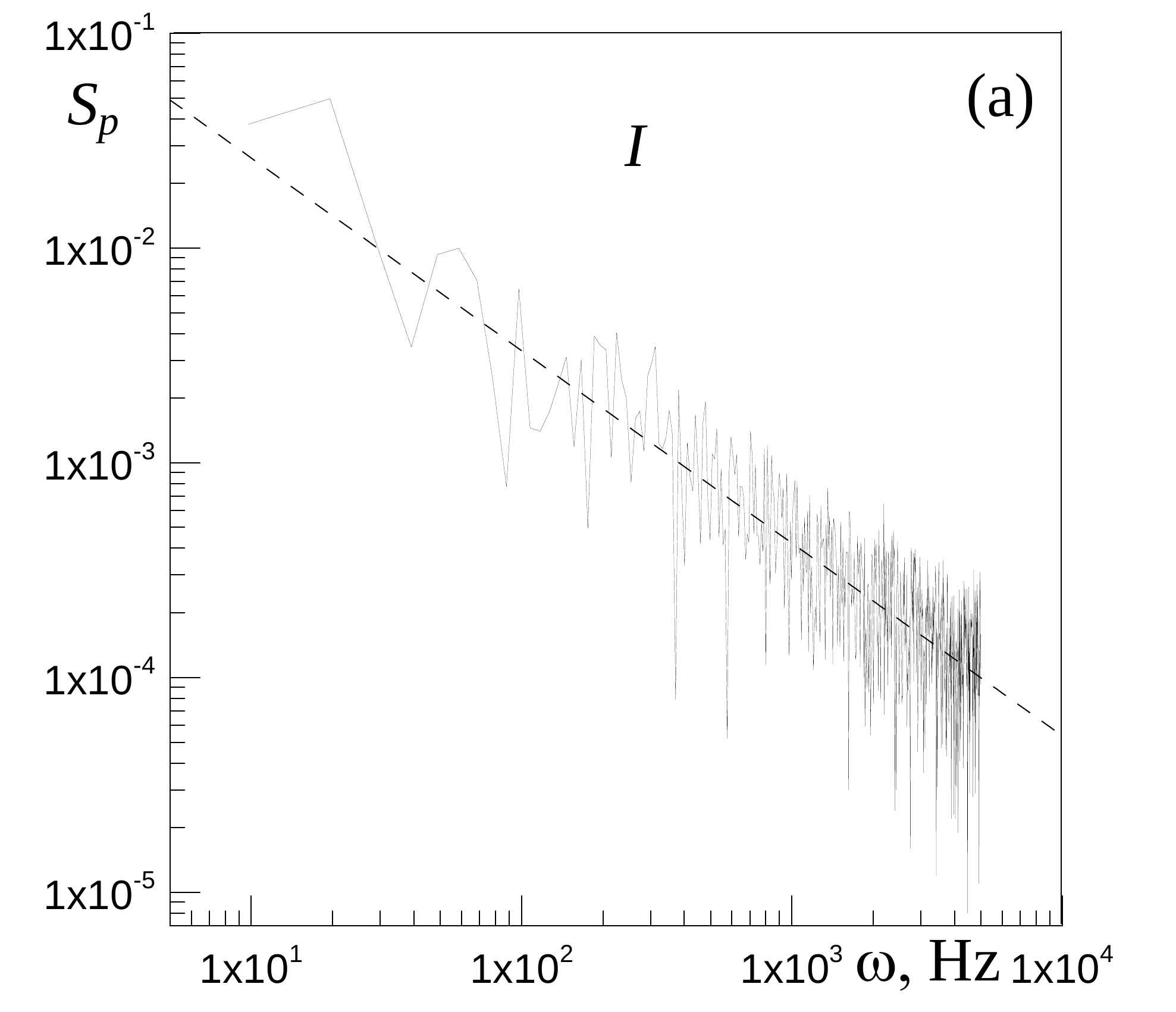} \includegraphics[width=0.5\textwidth]{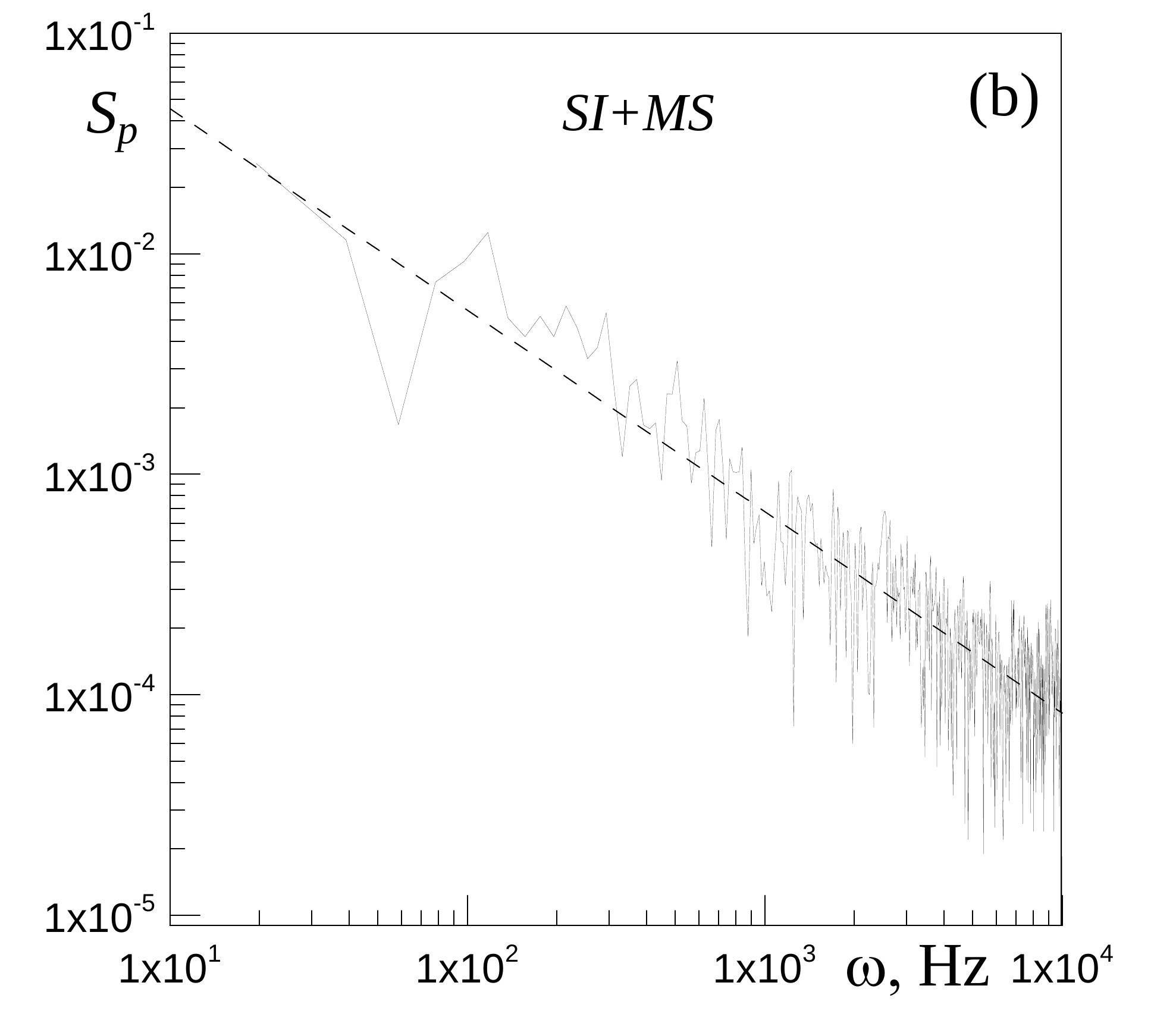}
}
\caption{Power spectrum of the signal $S_p(\omega)$ (in conventional units) for the $I$ (a) and $SI+MS$ (b) friction modes, which are depicted in figure~\ref{fig1} at the sampling rate $\omega_n=1/\triangle t\:\textrm{Hz}$ and the $1024$ frequency bands (Fourier lines). Dashed lines are determined by $S_p(\omega) \propto 1/\omega^{0.9}$ (a) and $S_p(\omega) \propto 1/\omega^{0.91}$ (b).}
\label{fig2}
\end{figure}

The results of spectral analysis of the time dependencies of the friction force $F(t)$ are presented in figure~\ref{fig2}. In particular, let us test the evolutionary dependencies for the presence of harmonic components that can be formed due to the influence of additive noncorrelated noise. The spectrum determination procedure is reduced to the application of the Fourier transform \cite{press2007numerical,Khom_PhysRevE_19} that is determined by
\begin{equation}\label{int_FFT_32}
 S_p = \frac{1}{P} \int_0^P f(t)\exp(-\ri\omega t)\rd t,
\end{equation}
where $P$ is function $f(t)$ period, $\omega=2\piup/P$ is fundamental frequency.
The calculation of the coefficients $S_p$ is usually carried out by using the fast Fourier transform method, since it is a recursive algorithm, which allows us to significantly save time during each iteration calculus~\cite{press2007numerical}.

This paragraph shortly describes the stages on what the algorithm of the fast Fourier transform is based. At first, we need to convert the problem into a discrete form. To do this, the discrete instants of time $t_n=n\cdot\triangle t$ are introduced, where $\triangle t$ is the sampling period. Further, we calculate the discrete values of the function at these instants of time $x_n=f(n\cdot\triangle t)$. In this case, the total period of the function is given by the multiplication of the total number of points and the sampling period $P=N\cdot\triangle t$. Correspondingly, the sampling rate of the signal equals $\omega=2\piup/P=2\piup/(N\cdot\triangle t)$. As a result, equation (\ref{int_FFT_32}) leads to the definition of a discrete Fourier transform for a one-dimensional array $x_n$ of length $N$
\begin{equation}\label{S_FFT_33}
 S_p = \frac{1}{N} \sum_{n=0}^{N-1} x_n\exp\left( \frac{-2\piup \ri n k}{N} \right),
\end{equation}
where $k=0,\:...,\:N-1$ is the index of discrete Fourier transform over the frequency \cite{press2007numerical}.

The time dependencies of the friction force (figure~\ref{fig1}) are analyzed on the basis of the algorithm of the fast Fourier transform \cite{press2007numerical}, which is described above. It calculates the spectrum of $F(t)$ evolution using (\ref{S_FFT_33}). Thus, the function~(\ref{S_FFT_33}) computes a vector of complex numbers. In fact, the modules of these values represent the amplitudes of the corresponding frequencies of the harmonic components of the signal, while the arguments give their initial phases. The entire spectrum equals a positive part with a doubled amplitude. Therefore, negative frequencies are neglected since a mirror image has no meaning. After normalization, we get the power spectrum $S_p(\omega)$ (figure~\ref{fig2}). In the friction force time series, there are no expressed (selected) frequencies of their regular (periodic) components, since $S_p(\omega)$ has no maxima. Let us note that $F(t)$ series are built using  iterations (\ref{iter}) at $P=100$, $\triangle t=10^{-4}$ ($I$), $P=50$, $\triangle t=5\cdot 10^{-5}$ ($SI+MS$), $N=10^6$.

It is apparent from figure~\ref{fig2} that the power spectrum $S_p(\omega)$ is mainly the same for both friction modes, i.e., the noise intensities indicated in figure~\ref{fig1}. Spectral densities descent monotonously in  logarithmic axes with an increase in the frequency. This implies that fluctuations $I_a(\varepsilon)$ are more powerful at low $\omega$. Correspondingly, the approximating dashed lines $S_p(\omega) = 0.21/\omega^{0.9}$ (see figure~\ref{fig2}a) and $S_p(\omega)= 0.37/\omega^{0.91}$ (see figure~\ref{fig2}b), also decrease with the ascent of frequency. The spectrum (energy) of fluctuations of the evolutionary variables is inversely proportional to the frequency for both ice rubbing modes. It means that various time correlations are present. This behavior is contrary to the properties of white noise, because in the case of white noise, the spectral density of the random signal assumes a constant value $S_p(\omega)={\rm const}$. It corresponds to the range of all possible frequencies, i.e., the signal power at all frequencies is the same. Consequently, there are no correlations in the system~\cite{gardiner2009}.

It is commonly recognized that natural systems, for which the phase transitions are inherent, have stochastic (fluctuation) processes with spectral density $\propto \omega^{-1}$ known as $1/\omega$, flicker or fractional (fractal) noise~\cite{gardiner2009,Ovchinnikov_2016,jetph_14}. Particularly, such a pattern takes place in many nonequilibrium systems that are characterized by Langevin stochastic source \cite{Shashkov_PhysRevE_17,MUDROCK20111093,Zakharov_NAP2018}. Thus, the spectrum cut-off is realized owing to  the nonlinear form of Langevin equation~(\ref{VI.5_17a}), describing mutually coordinated behavior of $\varepsilon, \sigma, T$, that counteracts the increase in frequency. Thus, white noise $\xi(t)$, which is typical of most natural systems, is transformed into a colored (correlated) one. The power exponents $\alpha=0.9$ and $\alpha=0.91$, which fit the graphs in figure~\ref{fig2}, demonstrate the presence of a ``pink'' noise. The noise strongly influences the pattern of a premelted film on the surface of ice, since it is very thin ($\sim$ 10 nm). However, similar spectral densities of the signal have been obtained for the refinement of grain structures during severe plastic deformation of volume specimens without spatial restrictions \cite{Khom_PhysRevE_19}.

As it is known, ``pink'' noise has a power spectrum fixed by expression $S_p(\omega) \propto 1 /\omega^\alpha$, $0<\alpha<2$. These fluctuations are widespread in nature and intermediate between ``white'' with a spectrum $S_p(\omega) \propto 1 /\omega^0$ and ``brown'' or ``red'' with $S_p(\omega) \propto 1 /\omega^2$ \cite{gardiner2009,Ovchinnikov_2016,Shashkov_PhysRevE_17,MUDROCK20111093,Zakharov_NAP2018,RCR_14,Gonch_APP_17,Yu_Bad_JNEP_12,Stef_PSS_16,Khomenko_NAP2017}. Besides, ``pink'' noise is a process with memory similar to the ``red'' one, i.e., the prehistory of the evolution of a casual process is considered. Contrary to that, ``white'' noise has no memory (correlations). We can conclude that the fluctuational spectrum at ice friction represents $1 /\omega^\alpha$ or ``pink'' noise, which displays the correlated fluctuations. The power-law relationships for approximations of spectral power density $S_p(\omega)$ confirm the self-affinity of friction force time series \cite{Persson2014}.

\begin{figure}[htbp]
\centerline{
\includegraphics[width=0.5\textwidth]{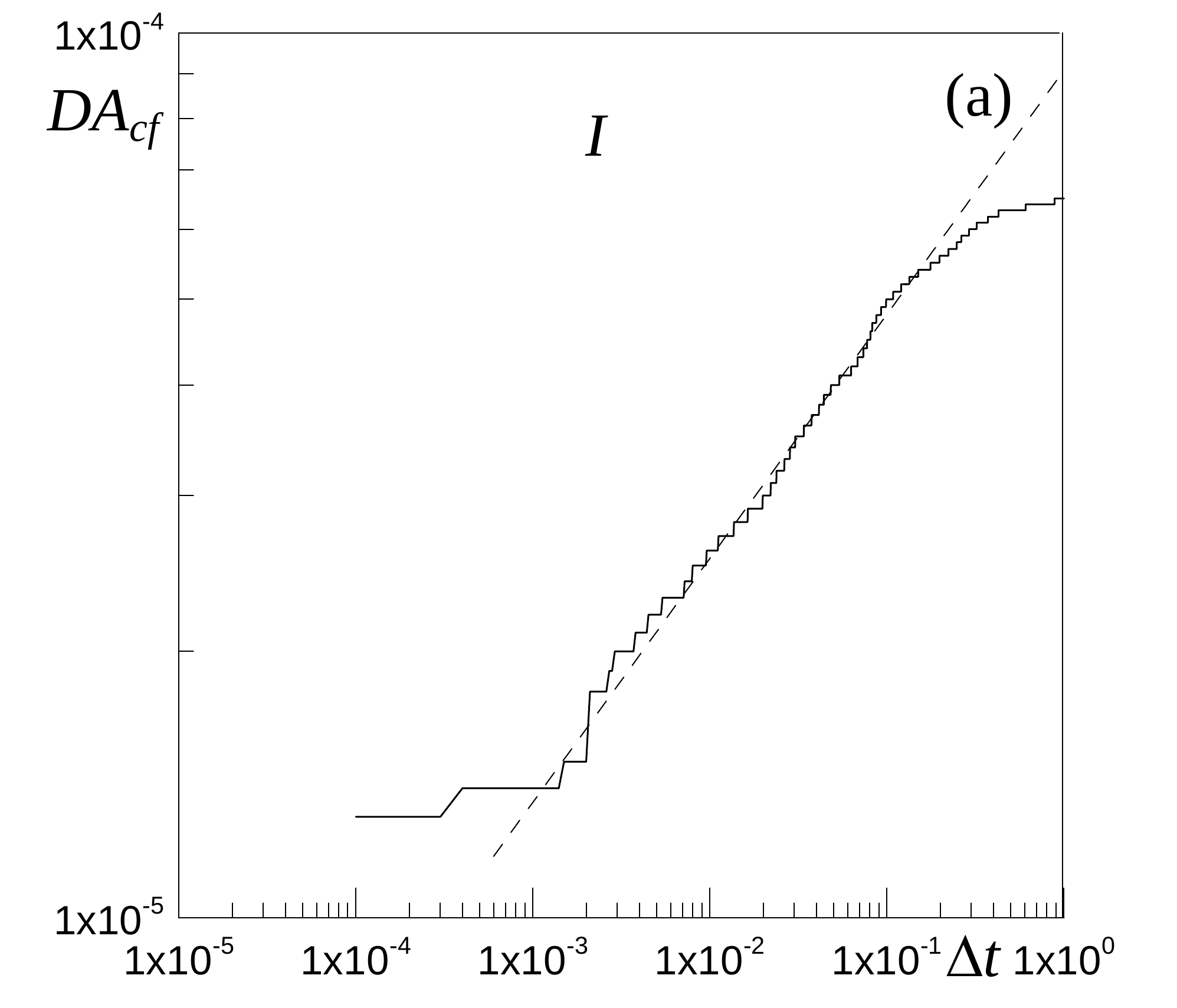} \includegraphics[width=0.5\textwidth]{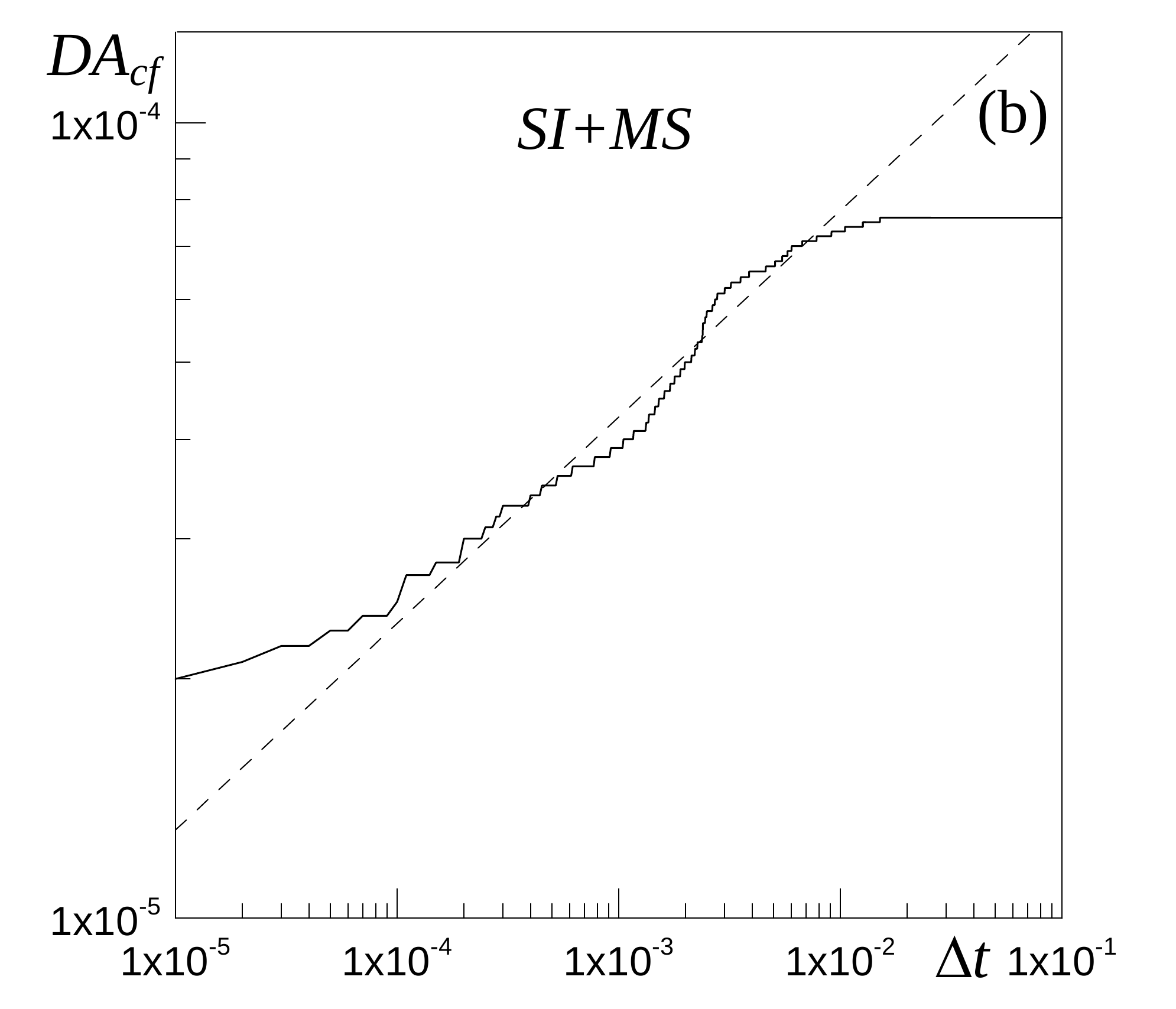}
}
\caption{Friction force-difference ACF $DA_{cf}(\triangle t)$ (in conventional units) for the $I$ (a) and $SI+MS$ (b) friction modes, which are represented in figures~\ref{fig1}~and~\ref{fig2}. Dashed lines are defined by $DA_{cf}(\triangle t) \propto \triangle t^{0.28}$ (a) and $DA_{cf}(\triangle t) \propto \triangle t^{0.26}$ (b).}
\label{fig3}
\end{figure}

Further, we construct a friction force-difference autocorrelation function (ACF) for time series analysis,  using the formula \cite{WANG_TL_2018}:
\begin{equation}\label{dacf}
DA_{cf}(\triangle t)=\frac{1}{2}\left\langle  \left\{F(t)-F(t+\triangle t)\right\}^2\right\rangle_t.
\end{equation}
Here, $F(t)$ is the friction force as a function of time, while $\langle \ldots \rangle_t$ denotes a time average value taken either over a fairly large range or over a periodically repeated regions. The coefficient 1/2 is introduced into the formula (\ref{dacf}) in order to convert $DA_{cf}(\triangle t)$ into a scaling form for the $F(t)$ variance that gives the mean-square $F(t)$ value at large $\triangle t$. Using a difference ACF permits to detect the self-affine properties of friction force time series additionally to an  ordinary fast Fourier transform. Figure~\ref{fig3} shows the power type of the difference ACF. It is important that the graphs are constructed in logarithmic coordinates, and this in turn allows us to trace the linear approximation. Hence, the power-law relationship for the difference ACF, which directly indicates $F(t)$ self-similarity. Besides, the values of power exponent smaller than $1$ are typical of the ice surface since it may have large random mean square slopes, i.e., fragility, and may be flattened under the external action \cite{Persson2014}.

Let us conduct a supplemental analysis of the random fluctuations of the rubbing force $F(t)$ by computing an ACF~\cite{WANG_TL_2018,Time_Series_An_2015}
\begin{equation}\label{S_ACF_b_34}
A_{cf}(\tau) = \kappa^2 - DA_{cf}(\tau),
\end{equation}
where $\kappa^2$ is the variance. The above formula shows the relationship between two random processes shifted relative to each other by several time $\tau$ (a signal delay). Autocorrelation function characterizes the complex oscillations, permitting to find their periodic components. The study of the extreme values of the obtained correlograms allows us to define the correlation times. It is supposed that the ACF is periodic, if the generic function is periodic~\cite{Time_Series_An_2015}.
For a discrete time series with mathematical expectation~$\lambda$, autocorrelation dependence is computed with the help of equation~\cite{Time_Series_An_2015}
\begin{equation}\label{S_ACF_35}
 A_{cf}(\tau) = \frac{1}{N\kappa^4}\sum_{t=1}^{N-\tau} (F_t-\lambda)(F_{t-\tau}-\lambda),
\end{equation}
where $N$ is the number of points in time series, $\tau$ is the lag (delay) time.

\begin{figure}[htbp]
\centerline{\includegraphics[width=0.45\textwidth]{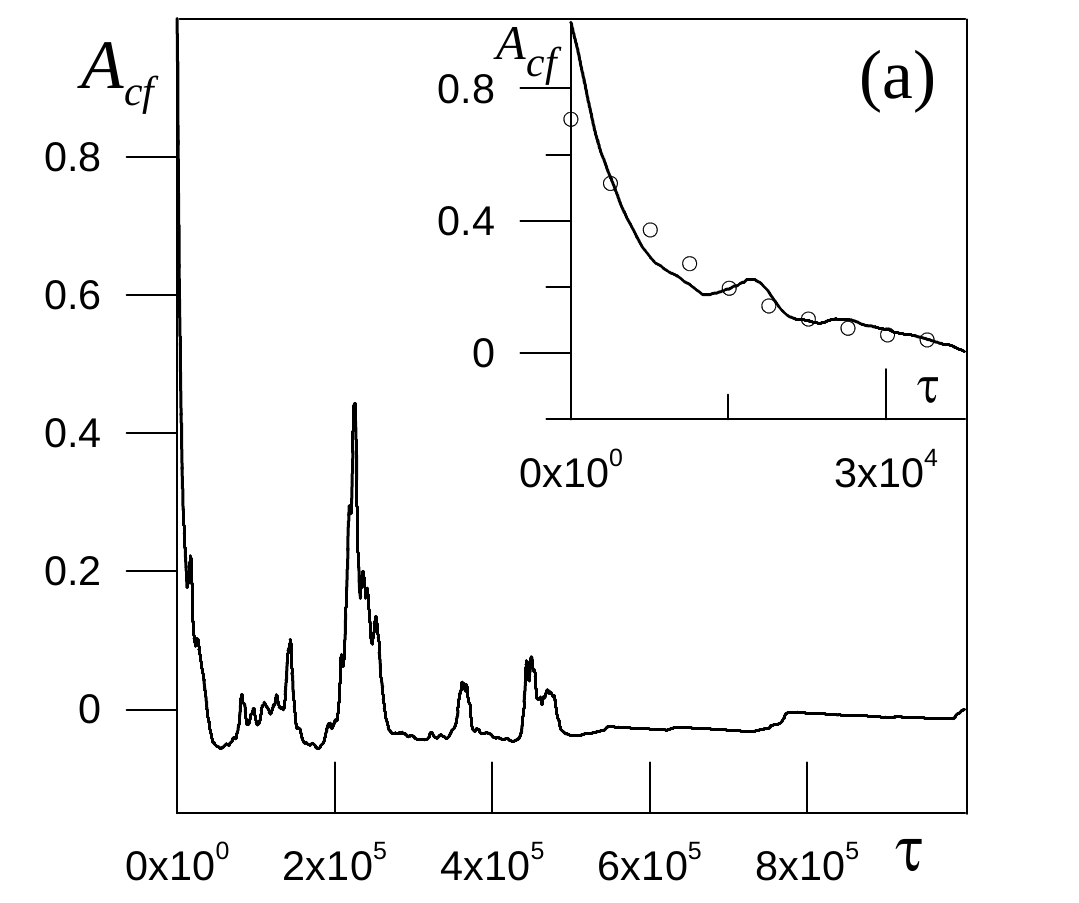} \includegraphics[width=0.45\textwidth]{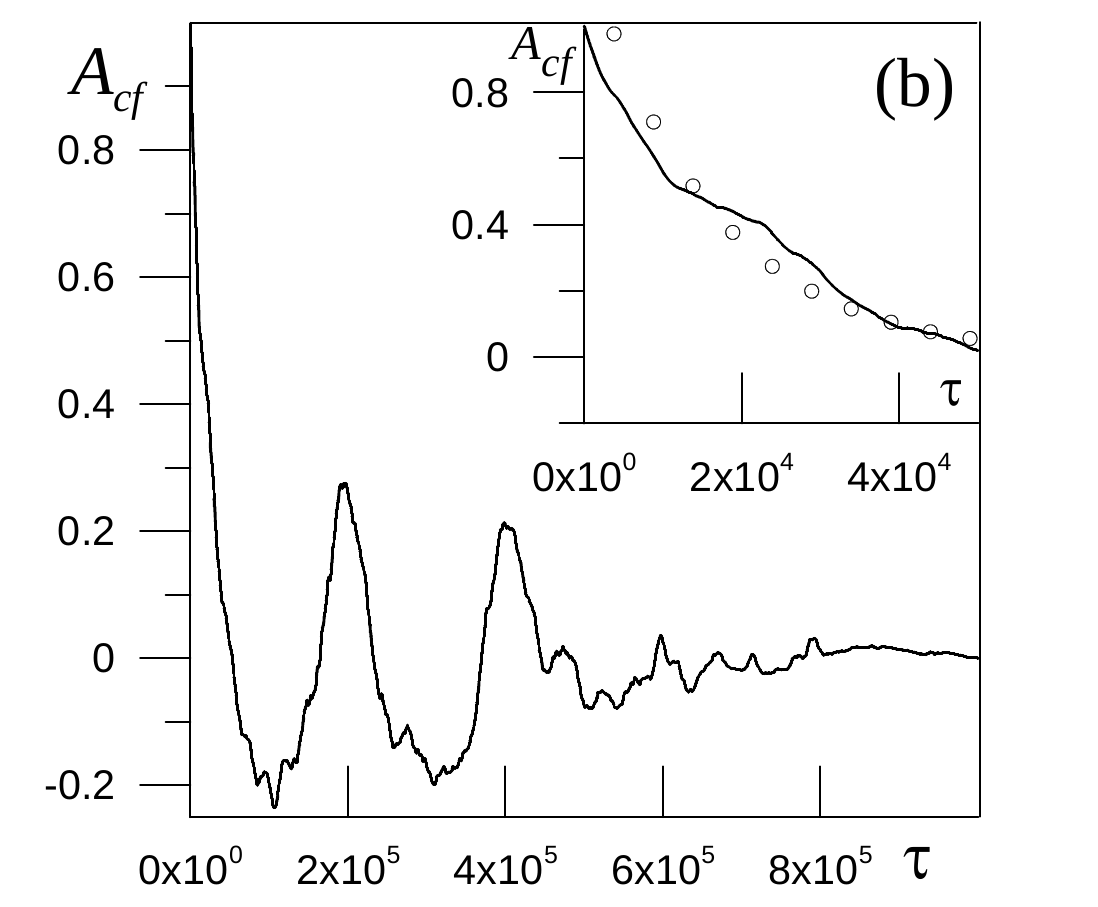}}
\caption{Autocorrelation functions $A_{cf}(\tau )$~(\ref{S_ACF_35}) of time series. The correlations, depicted in figures (a) and (b), are built for the time dependencies of the friction force at the conditions of figures~\ref{fig1}--\ref{fig3} and values $P=100$, $\triangle t=10^{-4}$ ($I$), $P=50$, $\triangle t=5\cdot 10^{-5}$ ($SI+MS$), $N=10^6$. Curves with circles are determined by equations $A_{cf}(\tau )= 0.707\exp(-8.514\cdot 10^{-5} \tau )$ (a) and $A_{cf}(\tau )= 1.239\exp(-6.323\cdot 10^{-5}\tau)$ (b).}
\label{fig4}
\end{figure}

Figure~\ref{fig4} demonstrates the ACFs of the basic evolutionary dependence and the shifted one by an amount~$\tau$ (shift value) in time, computed by the formula~(\ref{S_ACF_35}). Figures~\ref{fig4}a~and~\ref{fig4}b are calculated for the corresponding time series of the friction force in figure~\ref{fig1}. It is revealed that at the beginning, the ACFs decrease exponentially to zero (see inset) and, then, present damped oscillations around a certain value till approximately $5\cdot 10^{5}$ and $8\cdot 10^{5}$ for $I$ and $SI+MS$ modes, respectively \cite{Khomenko2018_TL}. Subsequently, ACF becomes zero at larger $\tau$, as it should be for the stationary process. Thus, the periodical oscillations are absent. It is worth noting that in accordance with the Cheddok scale in the range $0 < A_{cf} < 0.3$, the connection strength between the time series is weak. Hence, the evolution of friction force and autocorrelation occur by casual manner. The distribution functions $P_a(\varepsilon)$ (\ref{a}) for friction modes $\it I$ ($I_T = 5$) and $\it SI+MS$ ($I_T = 55$) are described by a power law corresponding to self-affine regime \cite{Khomenko2018_TL}. This mode is restricted by conditions $\varepsilon \ll 1$ and $I_\varepsilon, I_\sigma \ll I_T$. Thus, the ACFs have been obtained for different intensities of thermal fluctuations.

As it is visible from figures, in both cases ACFs have the maximum at $\tau =0$. This implies that the correlation time equals zero as for absolutely random process (``white noise''), which has ACF in the form of the Dirac function. However, comparison of the insets in figures~\ref{fig4}a~and~\ref{fig4}b with the approximation of $\delta$-function \cite{Khomenko2018_TL,Khom_PhysRevE_19} shows that an autocorrelation takes place within a certain correlation time. Thus, in the studied system correlated (colored) noise is realized. The colored Gaussian noise $\zeta(t)$ can be obtained as the solution of equation~\cite{Risken_eng_1989}:
\begin{equation}\label{OY_C_noize_35}
\frac{\rd\zeta(t)}{\rd t}= -\frac{1}{\tau_\zeta}\zeta(t) + \frac{1}{\tau_\zeta}\xi(t),
\end{equation}
where $\xi(t)$ represents the Gaussian white noise with intensity $I$ and moments
\begin{equation}\label{WN_moment_36}
  \langle \xi(t)\rangle=0,\quad\langle \xi(t)\xi(t') \rangle= 2I \delta(t-t').
\end{equation}
The correlation function of the process $\zeta(t)$ can be written as
\begin{equation}\label{CN_moment_37}
\langle \zeta(t)\zeta(t') \rangle= \frac{I}{\tau_\zeta} \exp\left(-\frac{|t-t'|}{\tau_\zeta}\right),
\end{equation}
where $\tau_\zeta$ is autocorrelation time.

The curves with circles (approximations) in the insets of figures~\ref{fig4}a~and~\ref{fig4}b are determined respectively by: $A_{cf}(\tau )= 0.707\exp(-8.514\cdot 10^{-5} \tau )$ and $A_{cf}(\tau )= 1.239\exp(-6.323\cdot 10^{-5}\tau)$. Thus, the evolutionary dependencies of the friction force $F(t)$ demonstrated in figure~\ref{fig1} have autocorrelation times $\tau \equiv 1/8.514\cdot 10^{-5}=1.175\cdot 10^{4}$ and $\tau \equiv 1/6.323\cdot 10^{-5}=1.582\cdot 10^{4}$. The given values imply that in the course of evolution the rubbing force preferably conserves the trend during a certain time $\tau$. Dependently on the direction, the correlation can be either positive (direct) or negative (reverse), for example, the increasing and decreasing parts of the dependence. At a positive correlation, the higher values of one feature correspond to the higher values of another one, but the lower values of one feature meet the lower values of another one. The connection of the present situation with the history of the process is more expressed at larger autocorrelation time $\tau$. Physically, we can predict which value of $F(t)$ will be there in the future. That is, there appears a possibility of predicting the ice friction type.

In the case of evolution of the friction force according to figure~\ref{fig4}b, it is assumed that autocorrelation time $\tau \equiv 1.582\cdot 10^{4}$ is  ``effective'', since the ice surface undergoes  repeated structural and kinetic transformations. These transitions consist in the variations between two ice rubbing modes (phases) with fixed $F(t)$ values \cite{Pers_JCP_18,TL_Pers_2016,Pers_15,Sazaki24012012}. The presence of two stable stationary modes in the region $SI+MS$  means that under the same conditions during friction, both these stationary modes can be realized simultaneously, but in different areas of the contact with the probability displayed in figure~\ref{fig1}. Moreover, due to the fluctuation nature of the process, these areas can randomly transit from one stationary state to another and vice versa. It should be noted that the formation of two-mode friction force distributions is actually observed in experiments \cite{TL_Pers_2016,Pers_15,Sazaki24012012}. In addition, one of the authors proposed a theory of the formation of a two-mode rubbing force distribution, based on a stochastic approach \cite{Khomenko2017_TL,Khomenko2018_TL}. In these models, small and large friction were formally considered as different types of surface structures.

It is supposed that, under certain conditions and for a determined time $\tau$, the ice slides most preferably in a steady state. According to the probability densities of friction force realizations (right-hand side of figure~\ref{fig1}), it implies that the ice surface layer has different composition. Thus, it is possible to provide the generation of a preferred stable structural state with a necessary rubbing at subsequent times. So, the above analysis shows that at ice friction there is a memory of the previous ice stable states during time $\tau$. Therefore, we can conclude that the magnitude of autocorrelation time influences the deviation of the scope of rubbing force from the average value. Consequently, the $F(t)$, in the formed ice structure, can take different values under the same conditions. On the other hand, at larger $\tau$, more time is needed to obtain the desired outcome. Thus, our study can be useful for applications for finding the frictional conditions to reach the sought result, that is stable ice rubbing mode.

\section{Summary}\label{sec:level4}

Using the Langevin equation, the time series of friction force are calculated for the following modes numerically: softened ice; crystalline ice; stable ice and metastable softening; metastable ice and stable softening. The analysis of time series for different modes of friction was performed using the fast Fourier transform. Oscillations were obtained with a spectral power density of the signal that is inversely proportional to the frequency and demonstrates the realization of $1 / \omega^\alpha$ or ``pink'' noise, which shows the presence of correlated fluctuations in the system. It is concluded that at $I$ and $SI+MS$ regimes, the time series are self-affine. The analysis of the friction force-difference autocorrelation function indicates the power type and, correspondingly, confirms the self-similarity of the series. The study of the autocorrelation function of the rubbing force random oscillations made it possible to determine the type of correlation and to identify the frequency characteristics of the process. It is revealed that the autocorrelation function is determined by an exponential dependence and demonstrates aperiodic behavior. Thus, these results can be useful in technical applications, since they allow us to predict the mode of friction during a certain time.

\section*{Acknowledgements}
This work is supported by the Ministry of Education and Science of Ukraine (Project ``Atomistic and nonlinear models of formation and friction of nanosystems'') and visitor grant of Forschungszentrum-J\"ulich, Germany. A. K. thanks Dr. Bo N. J.~Persson for hospitality during his stay in Forschungszentrum-J\"ulich.

%
%

\ukrainianpart

\title{Аналіз часових рядів сили тертя при самоподібному режимі розм'якшення поверхні льоду}
\author{О. В. Хоменко\refaddr{label1,label2}, Д. Т. Логвиненко\refaddr{label1}}
\addresses{
\addr{label1} Сумський державний університет, вул. Римського-Корсакова, 2, 40007 Суми, Україна
\addr{label2} Інститут Петера Грюнберга-1, Дослідницький центр Юліха, 52425, Юліх, Німеччина
}
%
%
%

\makeukrtitle

\begin{abstract}
\tolerance=3000%
В рамках реологічної моделі для апроксимації в'язкопружного середовища досліджується самоподібний режим розм'якшення льоду під час тертя. Вивчено різні режими тертя льоду, що визначаються утворенням поверхневого рідиноподібного шару. Проводиться аналіз часових рядів сили тертя, а саме Фур'є-аналіз, побудова автокореляційної та різницевої автокореляційної функцій. Спектральний степеневий закон виявляється для режимів кристалічного льоду, а також для суміші стійкого льоду та метастабільного розм'якшення. Доведено самоподібність та апериодичність відповідних часових рядів сили тертя.
\keywords розм'якшення поверхні льоду, в'язкопружне середовище, самоподібність, інтенсивність флуктуацій, автокореляційна функція, переривчасте тертя

\end{abstract}

\end{document}